\documentclass[11pt,fleqn]{article}
\usepackage{amsmath,amssymb,amsthm,enumerate}

\allowdisplaybreaks

\newcommand{\todo}[1][\null]{\ensuremath{\clubsuit}}
\newcommand{\noprint}[1]{}

\newcounter{mcasenum}

{\theoremstyle{definition}

}

\begin{document}

{\parindent=0mm

{\LARGE\bf Normalized Classes \\ of Generalized Burgers Equations\par}

\vspace{4mm}

{\large Oleksandr A. Pocheketa}

\vspace{2mm}

\emph{Institute of Mathematics of NAS of Ukraine,~3 Tereshchenkivska Str., 01601 Kyiv, Ukraine}

\vspace{0.2mm}

Email: pocheketa@yandex.ua

}

{\vspace{7mm}\par\noindent\hspace*{8mm}\parbox{144mm}{\small
A hierarchy of normalized classes of generalized Burgers equations
is studied.
The equivalence groupoids of these classes are computed.
The equivalence groupoids of classes of linearizable generalized Burgers equations
are related to those of the associated linear counterparts using the Hopf--Cole transformation.
}\par\vspace{2mm}}


\section{Introduction}

Consider some generalizations of the prominent Burgers equation
\begin{equation}\label{Pocheketa:BE}
u_t+uu_x+u_{xx}=0,
\end{equation}
which has been widely used as a~one-dimensional turbulence model~\cite{Pocheketa:burg48a}.
A review of its properties can be found in~\cite[Chapter~4]{Pocheketa:whit74book}.
The Burgers equation can be generalized in various ways.
The purpose of this paper is to study a~hierarchy of classes of generalized Burgers equations.
One may suppose that there are few normalized ones among them.
We show that the majority of naturally arising classes are normalized,
which considerably simplifies the solution of the group classification problems for these classes.
Namely, the problems reduce to subgroup analysis of the corresponding equivalence groups.

A class of differential equations is said to be \emph{normalized}
if its equivalence groupoid is generated by its equivalence group
\cite{Pocheketa:popo05,Pocheketa:popo06,Pocheketa:popo06k,Pocheketa:popo10a}.
The \emph{equivalence groupoid} of a~class of differential equations
is the set of admissible transformations in this class
with the natural groupoid structure, where the composition of mappings is the groupoid operation
\cite[p.~7]{Pocheketa:popo12a}.
An {\it admissible transformation} is a~triple of an initial equation, a~target equation and a~mapping between them.

The notion of normalized classes is quite natural and useful for applications.
For a~normalized class of differential equations
1)~its complete group classification coincides with its preliminary group classification and
2)~there are no additional equivalence transformations between cases of the classification list.
This notion can be weakened.
For example, weakly normalized classes
maintain the first of the aforementioned features but may lose the second,
and for semi-normalized classes the situation is opposite
(see \cite{Pocheketa:popo12a,Pocheketa:popo10a} for precise definitions).

Hierarchies of normalized subclasses arise
in the course of solving group classification problems.
Observe that a~single differential equation forms a~normalized class.
Any set of all possible equations with a~prescribed number of independent variables and a~fixed equation order
is a~normalized class likewise.

In order to prove the normalization property of a~class of differential equations
we compare its equivalence group with its equivalence groupoid.
Practically, a~class is normalized if there are no classifying conditions
among the determining equations for admissible transformations.
A classifying condition is, roughly speaking, a~determining equation
that simultaneously involves arbitrary elements of the class and parameters of admissible transformations
and leads to a~furcation while solving the determining equations. 

Section~2 is devoted to a~normalized superclass,
which contains all other classes under consideration.
In Section~3 we consider the relation between equivalence groupoids
of classes of linear (1+1)-dimensional evolution equations
and those of the associated classes of equations linearized by the Hopf--Cole transformation $u=2v_x/v$.
In Section~4 we consider classes of generalized Burgers equations with variable diffusion coefficients.
One of these classes is not normalized but it can be partitioned into two normalized subclasses.
Section~5 treats the classical Burgers equation as a~normalized class.

\section{Normalized superclass}

It is well known that the $t$-component of every point (or even contact) transformation
between any two fixed (1+1)-dimensional evolution equations depends only on~$t$
\cite{Pocheketa:king98a,Pocheketa:maga93a}.
Moreover, as proved in~\cite[Lemma~2]{Pocheketa:ivan10a},
any point transformation between two equations from the class
\begin{equation}\label{Pocheketa:supersuperclass}
u_t=F(t,x,u)u_{xx}+G(t,x,u,u_x)
\end{equation}
has the form
$\tilde t=T(t)$, $\tilde x=X(t,x)$, and $\tilde u=U(t,x,u)$ with $T_tX_xU_u\neq0$.
The coefficients~$F$ and~$G$ are
arbitrary smooth functions of their arguments with $F\neq0$.

This class is normalized in the usual sense~\cite{Pocheketa:ivan10a}, and
any contact transformation between equations from it
is generated by a~point transformation~\cite{Pocheketa:popo08c}.
However, class~\eqref{Pocheketa:supersuperclass} is too wide for the  generalized Burgers equations.
For our purpose it is more convenient to consider its subclass,
\begin{equation}\label{Pocheketa:superclass}
u_t+F(t,x,u)u_{xx}+H^1(t,x,u)u_x+H^0(t,x,u)=0,
\end{equation}
where the coefficients~$F$, $H^1$, and~$H^0$
are arbitrary smooth functions of their arguments with $F\neq0$.
This class is considered as the initial superclass for the present paper.
As it contains all subclasses to be studied,
any transformation between two fixed equations from each specified subclass
obeys the restrictions marked for class~\eqref{Pocheketa:supersuperclass}.

In order to find the general form of admissible transformations
for class~\eqref{Pocheketa:superclass},
we write an equation of this class in
tilded variables,
$\tilde u_{\tilde t}+\tilde F\tilde u_{\tilde x\tilde x}+\tilde{H^1}\tilde u_{\tilde x}+\tilde{H^0}=0$,
and replace $\tilde u_{\tilde t}$, $\tilde u_{\tilde x}$, and  $\tilde u_{\tilde x\tilde x}$
with their expressions in terms of untilded variables.
After restricting the result to the manifold defined by the initial equation
using the substitution $u_t=-Fu_{xx}-H^1u_x-H^0$,
we split it with respect to~$u_{xx}$ and~$u_x$ and obtain the determining equations for admissible transformations.
They imply
\begin{equation}\label{Pocheketa:superclass_AT}
\begin{split}
&\tilde t=T(t), \qquad \tilde x=X(t,x), \qquad \tilde u=U(t,x,u)=U^1(t,x)u+U^0(t,x), \\
&\tilde F=\frac{X_x^2}{T_t}F,\qquad
 \tilde H^1=\frac1{T_t}\left(X_xH^1+X_{xx}F-2X_x\frac{U^1_x}{U^1}F+X_t\right),\\
&\tilde H^0=U^1H^0+\frac{2U_xU^1_x}{T_tU^1}F-\frac1{T_t}\big(U_t+FU_{xx}+H^1U_x\big),
\end{split}
\end{equation}
where $T=T(t)$, $X=X(t,x)$, $U^1=U^1(t,x)$, and $U^0=U^0(t,x)$
are arbitrary smooth functions of their arguments
with $T_tX_xU^1\neq0$.
Note that we obtain no additional equations
(classifying conditions) on the arbitrary elements.
This means that all admissible transformations in this class are generated
by the transformations from the corresponding equivalence group,
so class~\eqref{Pocheketa:superclass} is normalized.

To derive admissible transformations of any subclass of~\eqref{Pocheketa:superclass}
it is sufficient to specify the arbitrary elements
$F$, $H^1$, $H^0$, $\tilde F$, $\tilde H^1$, and $\tilde H^0$.

\section{Linearizable generalized Burgers equations}

We relate the equivalence groupoids of the class of second-order linear evolution equations
and the 
class of linearizable generalized Burgers equations.
These equations have the forms
\begin{gather}
v_t+a(t,x)v_{xx}+b(t,x)v_x+c(t,x)v=0,\label{Pocheketa:Linearclass}\\
u_t+au_{xx}+(au+a_x+b)u_x+\frac12a_xu^2+b_xu+f=0,\label{Pocheketa:GLBEabc}
\end{gather}
respectively.
Here $a$, $b$, $c$ are smooth functions of $(t,x)$ with $a\neq0$, and $f=2c_x$.
Class~\eqref{Pocheketa:GLBEabc} is the widest class of differential equations
that can be linearized to linear equations of form~\eqref{Pocheketa:Linearclass}
by the Hopf--Cole transformation $u=2v_x/v$.
This linearization was implicitly presented in \cite[p.\,102, Exercise~3]{Pocheketa:fors1906book}.
Class~\eqref{Pocheketa:GLBEabc} is a~subclass of~\eqref{Pocheketa:superclass},
where the arbitrary elements are specified as
$F=a$, $H^1=au+a_x+b$, and $H^0=\frac12a_xu^2+b_xu+f$.
Substituting these and the corresponding tilded expressions into~\eqref{Pocheketa:superclass_AT}
and splitting the result with respect to~$u$,
we derive the general form of admissible transformations
between two equations from class~\eqref{Pocheketa:GLBEabc},
\begin{equation}\label{Pocheketa:GLBEabc_AT}
\begin{split}
&\tilde t=T(t), \qquad \tilde x=X(t,x), \qquad \tilde u=\frac1{X_x}u+U^0(t,x), \\
&\tilde a=\frac{X_x^2}{T_t}a,\qquad
\tilde b=\frac1{T_t}\big(X_xb+X_{xx}a-X_x^2U^0a+X_t\big),\\
&\tilde f=\frac{f}{T_t}-\frac{\big(X_xU^0b\big)_x}{T_t}+\frac{\big(X_xU^0\big)^2-2\big(X_xU^0\big)_x}{2T_t}a_x+{}\\
&\phantom{\tilde f=}+\frac{X_xU^0\big(X_xU^0\big)_x-\big(X_xU^0\big)_{xx}}{T_t}a-\frac{\big(X_xU^0\big)_t}{T_t},
\end{split}
\end{equation}
where $T=T(t)$, $X=X(t,x)$, and $U^0=U^0(t,x)$ are arbitrary smooth functions of their arguments
with $T_tX_x\neq0$.
There are no classifying conditions,
so, transformations~\eqref{Pocheketa:GLBEabc_AT} form the (usual) equivalence group,
and class~\eqref{Pocheketa:GLBEabc} is normalized (in the usual sense).

Arbitrary elements of class~\eqref{Pocheketa:GLBEabc} can be gauged to simple fixed values
by equivalence transformations.
At the first step we set $a=1$ using the transformation
\begin{equation*}
\tilde t=t\mathop{\rm sign}a(t,x), \qquad \tilde x=\int\frac{dx}{\sqrt{|a(t,x)|}}, \qquad \tilde u=u.
\end{equation*}
Thereby we obtain the class of equations of the general form
\begin{equation}\label{Pocheketa:GLBEbf}
u_t+u_{xx}+(u+b)u_x+b_xu+f=0,
\end{equation}
where $b=b(t,x)$ and $f=f(t,x)$ are arbitrary smooth functions.
The linear counterpart of~\eqref{Pocheketa:GLBEbf}
is
$v_t+v_{xx}+bv_x+\left(\frac12\int fdx\right)v=0$.

The equivalence group of class~\eqref{Pocheketa:GLBEbf} can be calculated directly
or by means of the substitution
$a=\tilde a=1$
into~\eqref{Pocheketa:GLBEabc_AT}.
It~consists of the transformations 
\begin{equation}\label{Pocheketa:GLBEbf_AT}
\begin{split}
&\tilde t=T(t), \qquad
 \tilde x=\varepsilon\left(\sqrt{T_t}x+X^0(t)\right), \qquad
 \tilde u=\varepsilon\left(\frac1{\sqrt{T_t}}u+U^0(t,x)\right), \\
&\tilde b=\varepsilon\left(\frac{b}{\sqrt{T_t}}+\frac{T_{tt}}{T_t^{3/2}}x+\frac{X^0_t}{\sqrt{T_t}}-U^0\right),\\
&\tilde f=\varepsilon\left(\frac{f}{T_t^{3/2}}-\frac{\big(U^0b\big)_x}{T_t}+\frac{U^0U^0_x}{\sqrt{T_t}}
 -\frac{U^0_t}{T_t}-\frac{U^0_{xx}}{T_t}-\frac{T_{tt}U^0}{2T_t^2}\right),
\end{split}
\end{equation}
where $T=T(t)$, $X^0=X^0(t)$, and $U^0=U^0(t,x)$ are arbitrary smooth functions
with $T_t>0$,
and the constant $\varepsilon$ takes the values $1$ and $-1$.
Class~\eqref{Pocheketa:GLBEbf} is normalized.

As the next step we set the arbitrary element~$b$ to zero by means of
the transformation
\begin{equation*}
\tilde t=t, \qquad \tilde x=x, \qquad \tilde u=u+b, \qquad \tilde f=f-b_t-bb_x-b_{xx},
\end{equation*}
which leads to the simplest reduced form for
linearizable generalized Burgers equations
containing the single arbitrary smooth function $f=f(t,x)$,
\begin{equation}\label{Pocheketa:GLBEf}
u_t+u_{xx}+uu_x+f=0.
\end{equation}
Substituting $b=\tilde b=0$ into~\eqref{Pocheketa:GLBEbf_AT}
we derive the general form of admissible transformations
between two equations of form~\eqref{Pocheketa:GLBEf},
\begin{gather*}
\tilde t=T(t), \qquad
\tilde x=\varepsilon\left(\sqrt{T_t}x+X^0(t)\right)\!, \qquad
\tilde u=\varepsilon\left(\frac1{\sqrt{T_t}}u+\frac{T_{tt}}{2T_t^{3/2}}x+\frac{X^0_t}{T_t}\right)\!, \\
\tilde f=\varepsilon\left(\frac{1}{T_t^{3/2}}f+\frac{3T_{tt}^2-2T_tT_{ttt}}{4T_t^{7/2}}x
+\frac{X^0_tT_{tt}-X^0_{tt}T_t}{T_t^3}\right),
\end{gather*}
where $T(t)$ is a~monotonically increasing smooth function,
$X^0(t)$ is an arbitrary smooth function,
and $\varepsilon=\pm1$.
Class~\eqref{Pocheketa:GLBEf} is normalized.
Its linear counterpart consists of equations of the form $v_t+v_{xx}+\left(\frac12\int fdx\right)v=0$.

Every equation from class~\eqref{Pocheketa:GLBEabc}
(resp.\ \eqref{Pocheketa:GLBEbf} or~\eqref{Pocheketa:GLBEf})
is connected with its linear counterpart via the Hopf--Cole transformation,
as well as the admissible transformations in any of these classes
are connected with
transformations in the corresponding linear classes.

Consider now the equivalence groupoid of the class of linear equations~\eqref{Pocheketa:Linearclass}.
It is determined by the transformations~\cite{Pocheketa:popo08a}
\begin{equation}\label{Pocheketa:Linearclass_AT}
\begin{split}
&\tilde t=T(t), \qquad
 \tilde x=X(t,x), \qquad \tilde v=V^1(t,x)v+V^0(t,x),\\
&\tilde a=\frac{X_x^2}{T_t}a,\qquad
 \tilde b=\frac1{T_t}\left(X_xb+X_{xx}a-\frac{2X_xV^1_x}{V^1}a+X_t\right),\\
&\tilde c=\frac1{T_t}\left(c-\frac{V^1_x}{V^1}b
 +\frac{2\big(V^1_x\big)^2-V^1V^1_{xx}}{\big(V^1\big)^2}a-\frac{V^1_t}{V^1}\right),
\end{split}
\end{equation}
where $T=T(t)$, $X=X(t,x)$, $V^1=V^1(t,x)$, and $V^0=V^0(t,x)$
are arbitrary smooth functions of their arguments
satisfying $T_tX_xV^1\ne0$
and the classifying condition
\begin{gather*}
\left(\frac{V^0}{V^1}\right)_t
+a\left(\frac{V^0}{V^1}\right)_{xx}
+b\left(\frac{V^0}{V^1}\right)_x
+c\frac{V^0}{V^1}=0.
\end{gather*}
This means that $V^0/V^1$ is a~solution of the initial equation~\eqref{Pocheketa:Linearclass}.
The equivalence group~$G^\sim$ of class~\eqref{Pocheketa:Linearclass}
consists of the transformations of form~\eqref{Pocheketa:Linearclass_AT}
with~$V^0=0$.
Class~\eqref{Pocheketa:Linearclass} is not normalized but semi-normalized
because
every transformation of form~\eqref{Pocheketa:Linearclass_AT}
is a~composition of the Lie symmetry transformation $\overline v=v+V^0/V^1$ of the initial equation
and an element of~$G^\sim$,
namely the transformation~\eqref{Pocheketa:Linearclass_AT} with~$V^0=0$.

A correspondence between the equivalence groupoids (resp.\ groups)
of classes~\eqref{Pocheketa:Linearclass}
and~\eqref{Pocheketa:GLBEabc}
can be established using the Hopf--Cole transformation.
Indeed,
\begin{gather*}
\tilde u=2\frac{\tilde v_{\tilde x}}{\tilde v}=\frac2{X_x}\frac{V^1v_x+V^1_xv+V^0_x}{V^1v+V^0}
=\frac1{X_x}\frac{\big(V^1u+2V^1_x\big)v+2V^0_x}{V^1v+V^0},
\end{gather*}
which writes in terms of $(t,x,u)$
only if~$V^0=0$.
The transformation component for~$u$ in this case is
\begin{gather*}
\tilde u=\frac1{X_x}u+\frac{2V^1_x}{X_xV^1}, \qquad \mbox{i.e.} \qquad U^0=\frac{2V^1_x}{X_xV^1}.
\end{gather*}
The constraint on~$V^0$ is related to
the general form of transformations
from the equivalence group of class~\eqref{Pocheketa:Linearclass}.
The admissible transformations with~$V^0\ne0$ in class~\eqref{Pocheketa:Linearclass}
have no counterparts in the equivalence groupoid of class~\eqref{Pocheketa:GLBEabc}.

Roughly speaking, the semi-normalization of class~\eqref{Pocheketa:Linearclass}
of linear equations
induces the normalization of class~\eqref{Pocheketa:GLBEabc} of linearizable equations.

\section{Generalized Burgers equations with arbitrary\\ diffusion coefficient}

Now we set $F=f(t,x)$, $H^1=u$, and $H^0=0$ in~\eqref{Pocheketa:superclass}.
This leads to the class of generalized Burgers equations
with an arbitrary nonvanishing smooth coefficient $f=f(t,x)$ of~$u_{xx}$,
\begin{equation}\label{Pocheketa:GBEtx}
u_t+uu_x+f(t,x)u_{xx}=0.
\end{equation}
Class~\eqref{Pocheketa:GBEtx} was considered, e.g., in~\cite{Pocheketa:king91c,Pocheketa:poch12a}.
Note that \cite{Pocheketa:king91c} is the first paper
where the exhaustive study of admissible transformations
of a~class of differential equations was carried out.
The equivalence group of class~\eqref{Pocheketa:GBEtx}
is finite dimensional and consists of the transformations
\begin{equation}\label{Pocheketa:GBEtx_group}
\begin{split}
&\widetilde t=\frac{\alpha t+\beta}{\gamma t+\delta},\quad
\widetilde x=\frac{\kappa x+\mu_1t+\mu_0}{\gamma t+\delta},\quad
\widetilde u=\frac{\kappa(\gamma t+\delta)u-\kappa\gamma x+\mu_1\delta-\mu_0\gamma}{\alpha\delta-\beta\gamma},\\
&\widetilde f=\frac{\kappa^2}{\alpha\delta-\beta\gamma}f,
\end{split}
\end{equation}
where the constant tuple
$(\alpha,\beta,\gamma,\delta,\kappa,\mu_0,\mu_1)$
is defined up to a~nonzero multiplier and satisfies the constraints
$\alpha\delta-\beta\gamma\neq0$ and $\kappa\neq0$.
The form of these transformations can be calculated directly
or by means of the substitutions
$F=f$, $\tilde F=\tilde f$, $H^1=u$, $\tilde H^1=\tilde u$, and $H^0=\tilde H^0=0$
into~\eqref{Pocheketa:superclass_AT}.
Since all transformations between any two fixed similar equations
from~\eqref{Pocheketa:GBEtx}
are exhausted by~\eqref{Pocheketa:GBEtx_group}, 
class~\eqref{Pocheketa:GBEtx} is normalized.

The class of equations of the form
\begin{equation}\label{Pocheketa:class_diffcoef}
u_t+uu_x+\big(f(t,x)u_x\big)_x=0
\end{equation}
with $f$ running through the set of nonvanishing smooth functions of~$(t,x)$ admits the transformations
\begin{equation}\label{Pocheketa:class_diffcoef_AT}
\begin{split}
&\tilde t=T(t), \qquad \tilde x=\varkappa\sqrt{|T_t|}x+X^0(t), \\
&\tilde u=\varkappa\frac{\sqrt{|T_t|}}{T_t}u
          +\varkappa\frac{T_{tt}\sqrt{|T_t|}}{2T_t^2}x + \frac{X^0}{T_t}, \qquad
 \tilde f=\varkappa^2f,
\end{split}
\end{equation}
where $\varkappa$ is an arbitrary nonzero constant
and the smooth functions $T$ and $X^0$ of~$t$ satisfy the equation
\begin{equation}\label{Pocheketa:class_diffcoef_eq}
\varkappa\mathop{\sqrt{|T_t|}}T_{tt}f_x+2T_tX_{tt}-2T_{tt}X_t=0.
\end{equation}
Unlike the previous classes, class~\eqref{Pocheketa:class_diffcoef} is not normalized.
At the same time, its subclass singled out by the inequality~$f_{xxx}\neq0$ is normalized.
In this case equation~\eqref{Pocheketa:class_diffcoef_eq} split with respect to~$f_x$
leads to the constraints~$X_{tx}=0$ and~$T_{tt}=0$.
Hence the associated equivalence groupoid is determined by the transformations
\begin{gather*}
\tilde t=c_1^2t+c_0, \quad
\tilde x=\varkappa c_1x+c_2t+c_3, \quad
\tilde u=\frac{\varkappa c_1u+c_2t+c_3}{c_1^2}, \quad
\tilde f=\varkappa^2f,
\end{gather*}
where $c_0$, $c_1$, $c_2$, $c_3$, and $\varkappa$ are arbitrary constants with $\varkappa c_1\ne0$,
which form the equivalence group of this subclass.

The complementary subclass, which is defined by the constraint~$f_{xxx}=0$, i.e., $f=f^2(t)x^2+f^1(t)x+f^0(t)$,
possesses a~wider equivalence groupoid.
Namely, all admissible transformations in this subclass are of form~\eqref{Pocheketa:class_diffcoef_AT},
where the parameter-functions $T=T(t)$ and $X^0=X^0(t)$
additionally satisfy the system of~ODEs
\begin{gather*}
4T_tT_{tt}f^2+2T_tT_{ttt}-3T_{tt}^{\,\,2}=0,\\
\frac\varkappa2\mathop{\sqrt{|T_t|}}T_{tt}f^1+T_tX^0_{tt}-T_{tt}X^0_t=0,
\end{gather*}
and $\varkappa$ is an arbitrary nonzero constant.
Although the general solution of this system is parameterized
by the arbitrary elements~$f^1$ and~$f^2$ in a~nonlocal way,
\begin{gather*}
T=\pm\int\left(C_2\int e^{-2\int f^2\,\mathrm dt}\,\mathrm dt+C_1\right)^{-2}\mathrm dt+C_0,\\
X^0=-\frac\varkappa2\int T_t\int \frac{\sqrt{|T_t|}T_{tt}}{T_t^{\,2}}f^1\,\mathrm dt\,\mathrm dt+C_3T+C_4,
\end{gather*}
the solution structure is the same for all values of the parameters.
In other words, the subclass singled out from class~\eqref{Pocheketa:class_diffcoef}
by the constraint~$f_{xxx}=0$ possesses a~nontrivial generalized extended equivalence group,
and it is normalized with respect to this group.
See, e.g.,
\cite{Pocheketa:ivan10a,Pocheketa:popo04a,Pocheketa:popo10a,Pocheketa:vane07a,Pocheketa:vane09a,Pocheketa:vane12a}
for the related definitions and other examples of generalized extended equivalence groups.

Note that the class of equations
$u_t+uu_x+f(t)u_{xx}=0$,
which differs from classes~\eqref{Pocheketa:GBEtx} and~\eqref{Pocheketa:class_diffcoef}
only in arguments of~$f$
and is the intersection of these classes,
is normalized with respect to
the equivalence group~\eqref{Pocheketa:GBEtx_group}
of the whole class~\eqref{Pocheketa:GBEtx}.
The group analysis of this class was performed in~\cite{Pocheketa:doyl90a,Pocheketa:wafo04d}.

\section{Classical Burgers equation}

To conclude, consider the class consisting of the single equation~\eqref{Pocheketa:BE}.
It is well known~\cite{Pocheketa:hopf50a,Pocheketa:cole51a}
that its linear counterpart is the heat equation $v_t+v_{xx}=0$.
The maximal Lie invariance algebra 
of the classical Burgers equation~\eqref{Pocheketa:BE}
is spanned by the vector fields~\cite{Pocheketa:katk65a}
\begin{equation*}
\begin{split}
\partial_t,\quad
2t\partial_t+x\partial_x-u\partial_u,\quad
t^2\partial_t+tx\partial_x+(x-ut)\partial_u,\quad
\partial_x,\quad
t\partial_x+\partial_u.
\end{split}
\end{equation*}
The complete point symmetry group 
of equation~\eqref{Pocheketa:BE} consists of the transformations
\begin{gather*}
\widetilde t=\frac{\alpha t+\beta}{\gamma t+\delta},\quad
\widetilde x=\frac{\kappa x+\mu_1t+\mu_0}{\gamma t+\delta},\quad
\widetilde u=\frac{\kappa(\gamma t+\delta)u-\kappa\gamma x+\mu_1\delta-\mu_0\gamma}{\alpha\delta-\beta\gamma},
\end{gather*}
where $(\alpha,\beta,\gamma,\delta,\kappa,\mu_0,\mu_1)$ is
an arbitrary set of constants defined up to a~nonzero multiplier, and
$\alpha\delta-\beta\gamma=\kappa^2>0$.
Up to composition with continuous point symmetries,
this group contains the single discrete symmetry $(t,x,u)\rightarrow(t,-x,-u)$.

\section{Conclusion}

This paper deals with a~hierarchy of normalized classes of generalized Burgers equations.
Due to the normalization property, the group classification for these classes
can be carried out using the algebraic method.
There are several examples of normalized classes the equivalence groups of which are finite dimensional,
which is an unexpected result.

It is important to emphasize the following phenomenon in the relationship
between the classes of linearizable generalized Burgers equations~\eqref{Pocheketa:GLBEabc}
and linear equations~\eqref{Pocheketa:Linearclass} as well as their subclasses via the Hopf--Cole transformation.
In view of the superposition principle for solutions of linear equations,
class~\eqref{Pocheketa:Linearclass} possesses
the wider set of admissible transformations than class~\eqref{Pocheketa:GLBEabc}.
Transformations associated with the linear superposition
depend on arbitrary elements of the corresponding initial equations.
This obstacle destroys the normalization property of class~\eqref{Pocheketa:Linearclass}, 
though this class is still semi-normalized in the usual sense.
At the same time, the linear superposition principle has no counterpart for the linearizable equations
among local transformations.
This is why class~\eqref{Pocheketa:GLBEabc} is normalized.

\subsection*{Acknowledgements}

The author is grateful to professor Roman~Popovych for his careful guidance and constructive help,
and to Olena Vaneeva and Vyacheslav Boyko for reading the manuscript and useful advice.

\end{document}